\documentclass[aps,prx,twocolumn,showpacs,10pt]{revtex4-1}
\usepackage{graphicx}
\usepackage{amssymb}
\usepackage{amsmath,bm}
\usepackage{comment}
\usepackage{footmisc}
\usepackage{color}
\usepackage{epsfig}
\def\de{\delta}
\def\De{\Delta}
\def\si{\sigma}
\def\dag{\dagger}
\pdfoutput=1

\begin{document}

\title{
Floquet Majorana and Para-Fermions in Driven Rashba Nanowires\\
}
 \author{Manisha Thakurathi}
\affiliation{Department of Physics, University of Basel,
            Klingelbergstrasse 82, CH-4056 Basel, Switzerland}   
\author{Daniel Loss}
\affiliation{Department of Physics, University of Basel,
            Klingelbergstrasse 82, CH-4056 Basel, Switzerland}
\author{Jelena Klinovaja}
\affiliation{Department of Physics, University of Basel,
            Klingelbergstrasse 82, CH-4056 Basel, Switzerland}

\begin{abstract}
We  study a periodically driven nanowire with Rashba-like  conduction and valence bands in the presence of a magnetic field. We identify topological regimes in which the system hosts zero-energy Majorana fermions. We further investigate the effect of strong electron-electron interactions that give rise to parafermion zero energy modes hosted at the nanowire ends. The first setup we consider allows for topological phases by applying only static magnetic fields without the need of superconductivity. The second setup involves both superconductivity and time-dependent magnetic fields and allows one to generate topological phases without fine-tuning of the chemical potential.
Promising candidate materials are graphene nanoribbons due to their intrinsic particle-hole symmetry.

\end{abstract}

\maketitle

\section{Introduction} 
Topological phases in condensed matter systems have been at the center of attention  over the past decade as they provide new venues for topological quantum computation. So far most of the studies on topological phases such as topological insulators \cite{bib:Kane2010,bib:Zhang2011,bib:Hankiewicz2013,bib:Pankratov1985,
bib:Volkov1987,bib:Zhang2006,bib:Zhang2007,bib:Zhang2009,Nagaosa_2009,bib:Moler2012,Amir_ext,
Leo_exp,Oreg}, Majorana fermions \cite{MF_Sato,MF_Sarma,MF_Oreg,alicea_majoranas_2010,MF_ee_Suhas,
potter_majoranas_2011,Klinovaja_CNT,Pascal,bilayer_MF_2012,Bena_MF,Rotating_field,
Ali,RKKY_Basel,RKKY_Simon,RKKY_Franz,mourik_signatures_2012,deng_observation_2012,
das_evidence_2012,Rokhinson,Goldhaber,marcus_MF,Ali_exp,Basel_exp}, and parafemions~\cite{PF_Linder,PF_Clarke,PF_Cheng,Ady_FMF,PF_Mong,vaezi_2,PFs_Loss,barkeshli_2} were focused on static systems. However, the dearth of naturally occurring topological materials is stimulating new proposals to engineer systems with topological phases.

External driving gives us a powerful tool to turn initially non-topological materials into  topological ones \cite{Fl_Nature_Linder}. This is a most promising approach for both condensed matter and cold atom fields. Recently, there have been several studies in which systems driven out of equilibrium give rise to a topological Floquet spectrum \cite{Ind,Platero_2013,Fl_Oka,Fl_Demler,Fl_Nature_Linder,Fl_PRB_Linder,Tanaka,Fl_Rudner,Fl_Liu,
Fl_Reynoso,Kat,Nagaosa,Fl_Grushin,Sen_MF1,Sen_MF2,Fl_Kitagawa,Fl_Rech,Fl_Katan,Fl_Foster,Fl_JK,Torres}. The existence of exotic edge modes have been demonstrated by direct observation in photonic crystals \cite{Fl_Kitagawa,Fl_Rech}.  The Floquet states have remarkably richer structure than its static counterparts.   There have been proposals on various novel phases of Floquet systems such as Floquet topological insulators \cite{Fl_Nature_Linder,Fl_Katan,Fl_JK,Kat},  Floquet topological superfluids \cite{Fl_Foster}, and Floquet Weyl semimetals \cite{Fl_JK,Nagaosa}. In this work, we explore one of such phases, namely,  Floquet fractional topological insulators which exhibit fractional excitations. This phase requires the presence of strong electron-electron interactions \cite{Ady_FMF,PFs_Loss,Fl_JK}, which is an interesting subject on its own in driven systems \cite{Sen,Polkovnikov}. 

\begin{figure}[t]
\centering
\includegraphics[width=0.9\columnwidth]{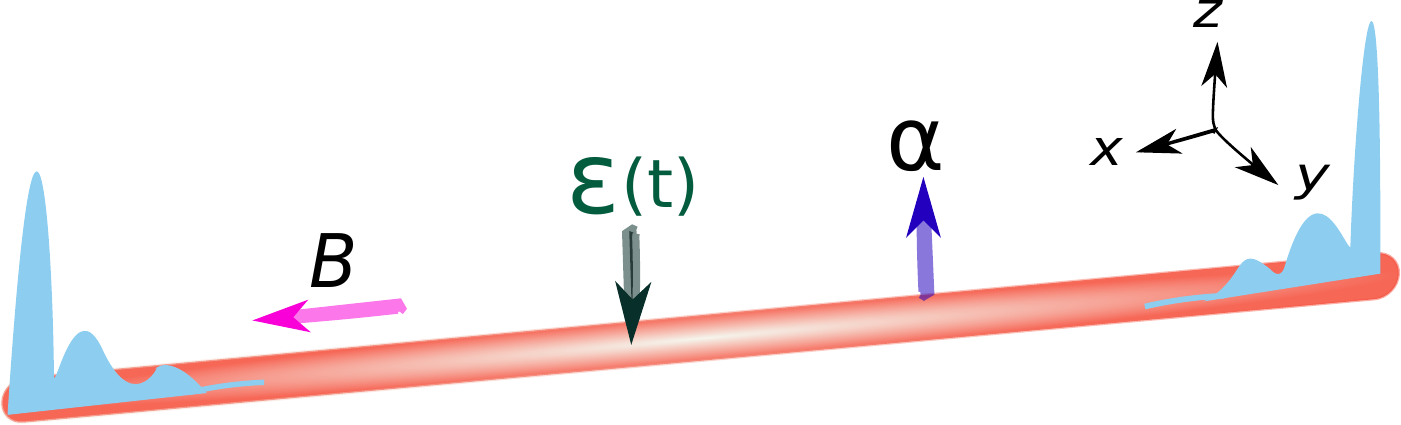}
\caption{One-dimensional Rashba nanowire (orange cylinder) with the SOI vector $\boldsymbol \alpha$ pointing in the $z$-direction is aligned along $x$-direction. The magnetic field $\bf B$ is chosen as well  to be in $x$-direction. A driving electric ac field $\boldsymbol{\mathcal{E}}(t)$ of frequency $\omega$, which matches the energy difference between conduction and valance bands and, thus, results in the coupling  of the  strength $t_F$ between bands, is applied transverse, say in $z$-direction. In the topological regime $\Delta_Z> t_F > 0$, the system hosts zero energy bound states (blue curves) at each wire end.}
\label{1D_one}
\end{figure}

In the first setup, we consider a Rashba nanowire (see Fig.~\ref{1D_one}) driven by an oscillating electric field $[\boldsymbol{\mathcal{E}}(t)]$ with frequency matching the energy difference between the conduction and valence bands. We note that our results are applicable to any single-channel system such as semiconducting nanowires, graphene nanoribbons, and nanotubes \cite{CNT,CNT2,CNT3,CNT4,gra,gra2,gra3,
gra4,gra5,sc,mos,mos2,mos3,mos4,mos5,mos6}. We show that the topological zero energy bound states localized at the nanowire ends can be realized by the mere presence of a uniform static magnetic field without any need of superconductivity. This proposal is attractive experimentally as it avoids the detrimental combination of magnetic fields and superconductivity. In the second setup, a one-band Rashba nanowire with proximity-induced superconductivity is subjected to a time-dependent magnetic field. This setup has an important advantage over those with  time-independent magnetic fields \cite{MF_Sarma,MF_Oreg} in that the chemical potential does not need to be tuned close to the spin-orbit energy.   For both setups, we find topological bound states in the fractional charge regime. 

These setups not only provide a proof-of-principle for fractional topological effects in Floquet systems but also show great promise to be experimentally  implemented
in realistic systems such as graphene nanoribbons. 

\section{Floquet Rashba nanowire in applied magnetic field}
\begin{figure}[b]
\includegraphics[width=1\columnwidth]{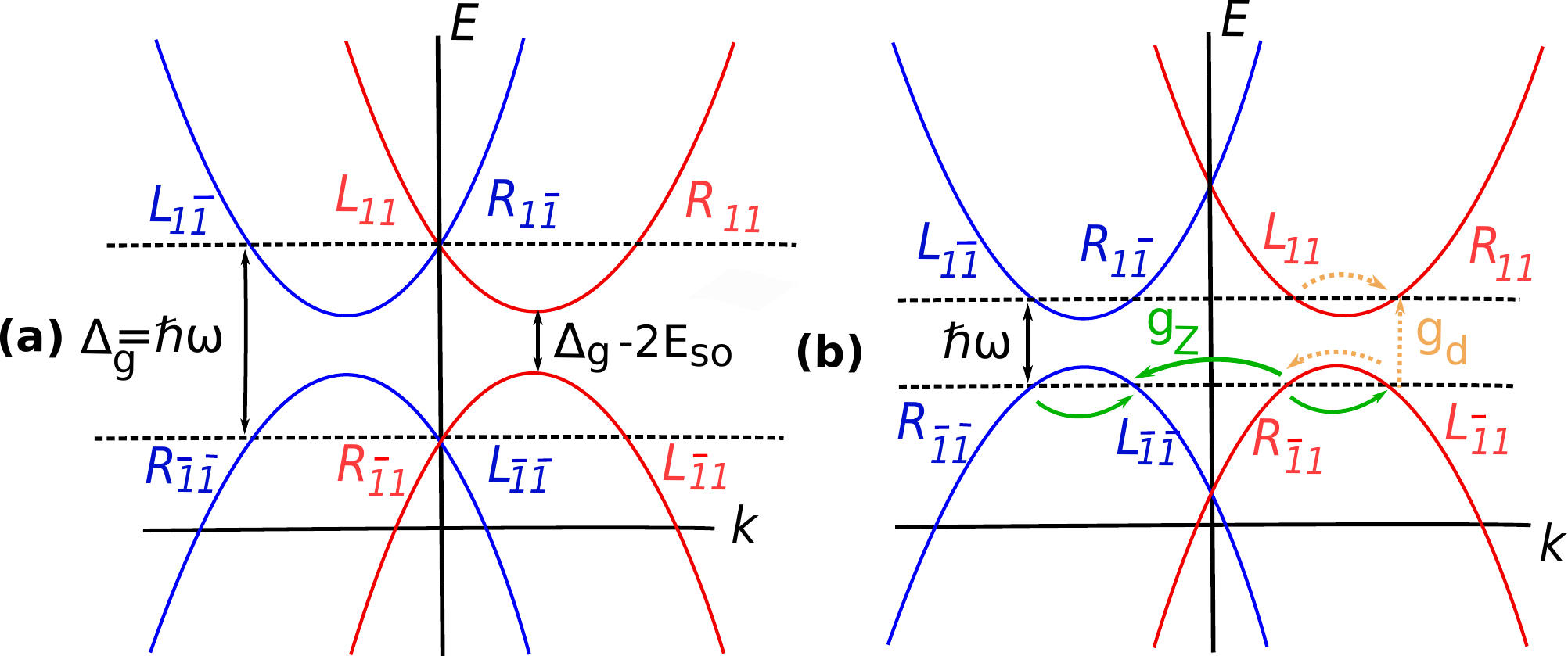}
\caption{The spectrum of the Rashba nanowire with the band gap of $\De_g-2 E_{so}$ separating the valence ($\eta=\bar  1$) and conduction ($\eta=1$) bands. The index $\sigma=1$ ($\sigma=\bar 1$) refers to the spin up (spin down) band shown in red (blue) color. 
The right- ($R_{\eta \sigma}$) and left-mover  ($L_{\eta \sigma}$) fields are introduced close to the Fermi level in the Floquet representation.
(a) The chemical potential $\mu$ is tuned close to the SOI energy $E_{so}$ and the driving frequency is chosen such that $\hbar \omega =\De_g$, resulting in resonant scattering between the two bands. 
If the Floquet amplitude $t_F$ is smaller than the Zeeman energy $\Delta_Z$, the system hosts MFs. (b) To obtain parafermions, we tune $\mu$ to $E_{so}/9$ and readjust the driving frequency to $\hbar \omega =\De_g - 16 E_{so}/9$. The leading term in the magnetic field $H_Z^{ee}$ (green arrows) involves two back-scattering events and opens a partial  gap in the spectrum only in the presence of strong electron-electron interactions. The driving term $H_d^{ee}$ (yellow arrows) commutes with $H_Z^{ee}$ and can be ordered simultaneously in the RG sense, leading to a fully gapped spectrum.
}
\label{Ek}
\end{figure}

We consider a one-dimensional Rashba nanowire (see Fig.~\ref{1D_one}) aligned along $x$- direction characterized by the spin-orbit interaction (SOI) vector $\boldsymbol \alpha$, which points perpendicular to the nanowire axis in the $z$-direction. The corresponding Hamiltonian  is given by
\begin{align}
H_0 =\sum_{\eta\sigma} \eta \Psi_{\eta \sigma }^\dagger   \left(- \frac{\hbar^2 \partial_x^2}{2 m_0 } + \alpha \sigma \partial_x  +\de_{1\eta}\De_g \right)\Psi_{\eta \sigma }. \label{H0}
\end{align}
Here, $m_0$ is the effective electron mass. The index $\eta=1$ ($\eta=\bar 1$) corresponds to the conduction (valance) band  and $\si=1$ ($\si=\bar1$) to spin up (down) states.  The fermion operator $\Psi_{\eta \sigma }(x)$ annihilates at position $x$ an electron from the $\eta$ band with spin $\sigma$. In the valence band, we initially tune the chemical potential $\mu$ close to the SOI energy $E_{so}=\hbar^2 k_{so}^2/2m_0$, where $k_{so}=m_0 \alpha/\hbar^2$ is the SOI wavevector. The gap between valence and conduction bands is $\Delta_g-2 E_{so}$, as shown in Fig. \ref{Ek}(a). 
A static and uniform magnetic field $\bf B$ is  applied perpendicular to the SOI vector (say, along $x$-direction) and results in the Zeeman term 

\begin{align}
H_Z=\De_Z \sum_{\eta \sigma \sigma'} \Psi_{\eta \sigma }^\dagger  (\sigma_x)_{\sigma\sigma'}\Psi_{\eta \sigma' }, \label{HZ}
\end{align}
where $\Delta_Z=g\mu_B B$ is the Zeeman energy with $g$ being the $g$-factor and $\mu_B$  the Bohr magneton.

Instead of making use of the standard  scheme  based on superconductivity \cite{MF_Sato,MF_Sarma,MF_Oreg}, we propose to drive the system across the bulk gap  by an oscillating electric field with frequency $\omega$. When the driving frequency matches the resonance energy, $\hbar \omega = \De_g$, a dynamical gap emerges in the system (playing the role of a superconducting gap).

\begin{figure}[t]
\begin{center}
\hspace*{-.5cm}
\includegraphics[width=3in,height=2in]{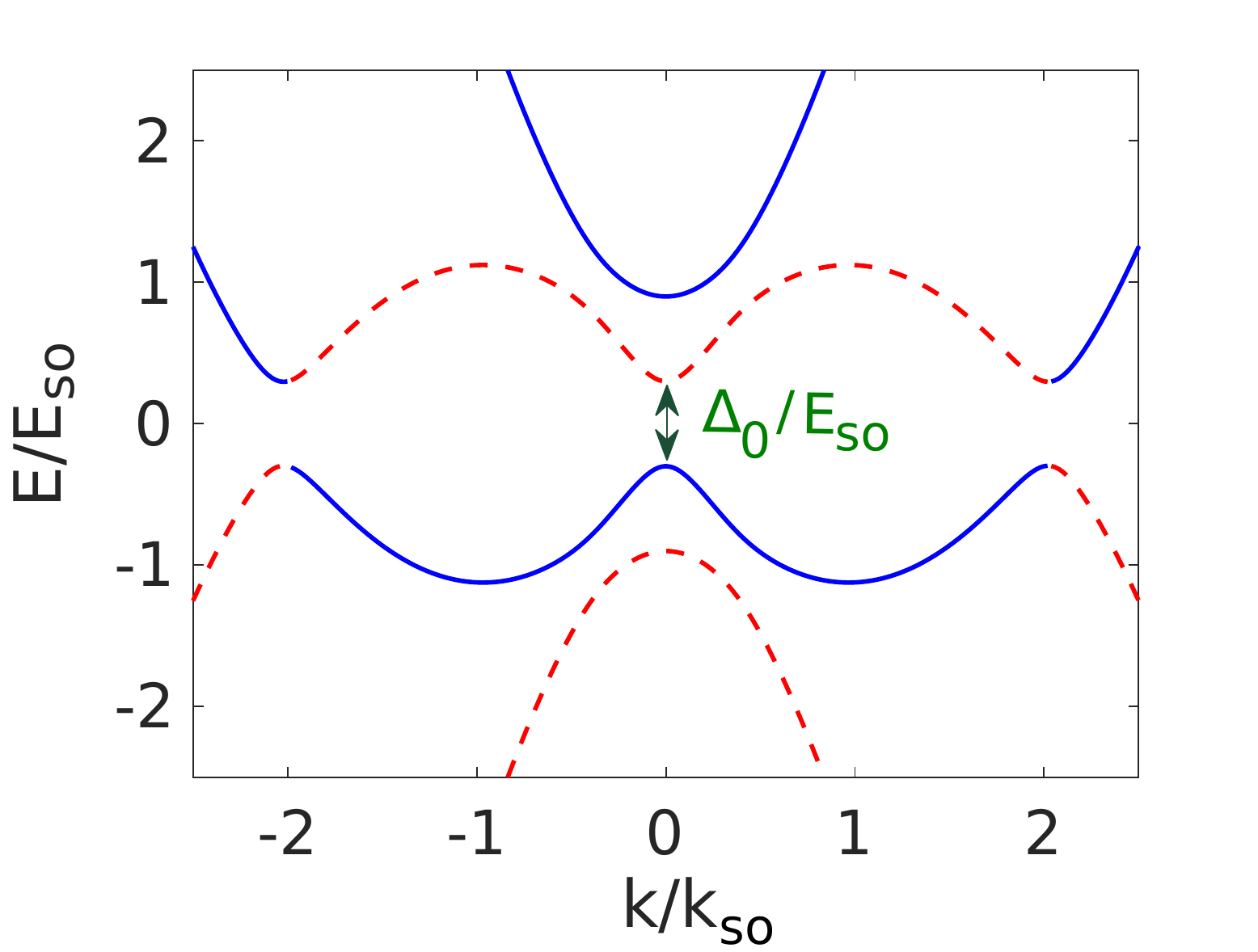}
\end{center}
\caption{Floquet spectrum [see Eq.~(\ref{spectrum1})] of  Rashba nanowire driven by electric fields  for  $\De_Z/E_{so}=0.6 $ and $t_F/E_{so}=0.3 $. The topological gap $\Delta_0=2 |\Delta_Z-t_F|$  defined at $k=0$ closes for $\Delta_Z=t_F$ signaling a topological phase transition. If $t_F>\Delta_Z$, there is one MF localized at each end of the nanowire.} 
\label{bulk}
\end{figure}

We work in the Floquet representation \cite{Fl_Oka,review}. To map a time-dependent problem into a stationary one, we replace the initial  time-dependent  periodic Hamiltonian $H(t)= H (t+T)$ by the Floquet Hamiltonian defined by $H_F = H (t) - i \hbar \partial_t$. The eigenstates of $H_F$ are given by the direct product  of the instantaneous  eigenstates $(|\nu\rangle\equiv| k \rangle \bigotimes |\eta \rangle \bigotimes |\sigma\rangle)$ and the set of periodic functions $e^{i n \omega t}$, where the integer  $n$ defines the $n^{\rm th}$ Floquet replica.  The matrix elements then become $\langle \nu_1 n_1|H_F|\nu_2,n_2\rangle=\langle \nu_1 |H|\nu_2\rangle + n_1 \hbar \omega \delta_{n_1 n_2} \delta_{\nu_1 \nu_2}$. We consider only the direct resonances between $n=0$ and $n=1$ involving single photon absorption/emission  processes and we work in first order approximation in the driving amplitude. The Floquet term, which couples conduction and valence bands, is given by $H_d =t_F \sum_{\eta \sigma} \Psi_{\eta \sigma }^\dagger \Psi_{\bar{\eta} \sigma}$, with
 the Floquet coupling amplitude $t_F = e \mathcal{E} d_{cv}/2 $ being proportional to the interband dipole term between conduction and valence band $(d_{cv})$ and to the amplitude of the applied electric field $\mathcal{E}$ {\cite{Fl_JK}}.
 Thus, in the basis $(\Psi_{11},\Psi_{1 \bar 1},\Psi_{\bar 11},\Psi_{\bar1 \bar1})$, the Floquet matrix assumes the form
\begin{align}
{\mathcal H}_F=\begin{pmatrix}
E_k+\alpha k& \Delta_Z& t_F&0 \\
\Delta_Z& E_k-\alpha k&0& t_F\\
t_F&0& -E_k-\alpha k& \Delta_Z \\
0&t_F&\Delta_Z&-E_k+\alpha k
\end{pmatrix},
\label{Hf}
\end{align}
where $E_k=\hbar^2 k^2/2m_0$. We note that $\Delta_g$ in the upper two diagonal elements is cancelled out by $\hbar\omega$.  The spectrum of ${\mathcal H}_F$
 [see Fig.(\ref{bulk})] consists of four branches, 
\begin{align}
E_{F\pm}^2= &\left(\frac{\hbar^2 k^2}{2m_0}\right)^2+ (\alpha k)^2 + \De_Z^2 +t_F^2 \nonumber \\  & \pm 2 \sqrt{\De_Z^2 t_F^2 + \left(\frac{\hbar^2 k^2}{2m_0}\right)^2 [(\alpha k)^2 + \De_Z^2]}.
\label{spectrum1}
\end{align}
The gap $\Delta_0=2 |\Delta_Z-t_F|$ at $k=0$ is zero only for $\Delta_Z=t_F$. At all other values of wavevector $k$, the gap in the Floquet spectrum is always finite. 
The closing of the gap $\Delta_0$ indicates two possible topological phase transition point with two phases characterized by $\Delta_Z<t_F$ and $\Delta_Z>t_F$.

Next, we identify the parameter regime in which the system is in the topological phase and hosts  Majorana fermion (MF) zero-energy modes localized at the wire ends. For simplification, we work in the regime of strong SOI and linearize the Hamiltonian ${\mathcal H}_F$ [see Eq. (\ref{Hf})] at the Fermi surface \cite{Composite,SOI_rote} by representing operators in terms of slowly-varying left ($L_{\eta \sigma}$) and right mover fields ($R_{\eta \sigma}$) defined around the Fermi points $k_F=\pm 2 k_{so}$ and $k_F=0$ (see Fig.~\ref{Ek}) as
\begin{align}
\Psi_{\eta \sigma} = R_{\eta \sigma} e^{i \sigma k_{so} (1+\eta \sigma)} + L _{\eta \sigma }  e^{i \sigma  k_{so} (1-\eta \sigma)}.
\end{align}
The effective Hamiltonian density $\mathcal H$ is written in terms of Pauli matrices 
in  the basis $(R_{11}, L_{11}, R_{1\bar 1}, L_{1 \bar 1}, R_{\bar 1 1}, L_{\bar 1 1}, R_{\bar 1\bar 1}, L_{\bar 1\bar 1})$ as
\begin{align}
\mathcal H = \hbar \upsilon_F \hat k \tau_3 + t_F \eta_1 \tau_1 + \De_Z ( \tau_1\sigma_1 + \eta_3 \tau_2 \sigma_2)/2,
\end{align}
where $\upsilon_F$ is the Fermi velocity and $\hat k$ the momentum operator with eigenvalue $k$. We note that the system is assumed to be in the weak driving regime  with $t_F \ll \De_g$. The Pauli matrices $\eta_i$ ($\sigma_i$) act in upper-lower (spin) spaces (subspaces) and $\tau_i$ act in right-left mover subspace. 

The corresponding Floquet spectrum is given by $E_{1,\pm}=\pm \sqrt{(\hbar \upsilon_F  k)^2+t_F^2}$ and $E_{2 \pm}^2= \sqrt{(\hbar \upsilon_F  k)^2+(t_F\pm \De_Z)^2}$,
where $E_{1,\pm}$ is twofold degenerate. If $\De_Z>t_F>0$, the system is in the topological phase and hosts one localized zero-energy state at each wire end. The corresponding wavefunction of the state localized at the left end  ($x=0$) is given  in the basis $(\Psi_{11},\Psi_{1 \bar 1},\Psi_{\bar 11},\Psi_{\bar1 \bar1})$ by
\begin{align}
&\Phi(x)=\begin{pmatrix}
f(x), i f^*(x), -i  f(x), - f^*(x) 
\end{pmatrix}^T, \\
&f(x)= e^{-x/\xi_t} e^{-2 i k_{so} x} - e^{-x/\xi_-},
\end{align}
with the localization lengths defined as $\xi_t=\hbar \upsilon_F/t_F$ and $\xi_-=\hbar \upsilon_F/(\De_Z-t_F)$.

\section{ Floquet Parafermions \label{FFM}}
The topological phases can also be realized in an interacting system, giving rise to fractional Floquet modes.  For example, if the chemical potential is moved down to $\mu_{1/3}=E_{so}/9$, such that the Fermi wavevectors are given by $\pm k_{so} (1\pm 1/3)$, the nanowire hosts parafermions as we show next. We note that we work in the high frequency limit meaning that the driving frequency $\omega$  is larger than any frequency associated with the internal dynamics of the system, and electron-electron interactions can be treated with standard bosonization techniques.

Similarly to the non-interacting model, we assume that the Zeeman field is the dominant term and drives the system into the topological phase.  The term, which conserves both spin and momentum and is lowest order in  $\De_{Z}$, is given by
\begin{align}
H_{Z}^{ee} =& g_Z  \Big[( R^\dagger_{1\bar  1}L_{11})( R^\dagger_{1\bar  1}L_{1\bar  1}) ( R^\dagger_{11}L_{11})\nonumber\\
&\hspace{15pt}+(R_{\bar 1 1}^\dagger L_{\bar 1\bar 1}) (R_{\bar 1 1}^\dagger L_{\bar 1 1})(R_{\bar 1\bar 1}^\dagger L_{\bar 1\bar 1}) + \text{H.c.} \Big],
\label{HZee}
\end{align}
where  $g_Z \propto \De_Z g_B^2$ and $g_B$ is the electron-electron back-scattering amplitude. This process involves the back-scattering of two electrons \cite{Ady_FMF,PFs_Loss}. Again, the frequency of the driving term matches the energy difference between the conduction and  valence bands, see Fig. \ref{Ek}(b). For weak driving, it is sufficient to include electron-electron interactions inside each of the two bands. The term, which commutes with $H_{Z}^{ee}$ and satisfies the momentum and energy conservation laws resulting in the dynamic gap, is written as
\begin{align}
H_{t}^{ee} =& g_d  \Big[( R^\dagger_{11}L_{\bar 1 1})
( R^\dagger_{11}L_{1 1})( R^\dagger_{\bar 1 1}L_{\bar 1 1})\nonumber\\
&\hspace{15pt}+(R_{\bar 1\bar 1}^\dagger L_{1 \bar  1}) (R_{\bar 1\bar 1}^\dagger L_{\bar 1 \bar1}) (R_{1 \bar  1}^\dagger L_{1 \bar  1})  + \text{H.c.} \Big],
\label{Hamflo}
\end{align}
where  $g_d \propto t_F g_B^2$. We assume that these two terms, Eqs.~(\ref{HZee}) and (\ref{Hamflo}), are relevant in the sense of the renormalization group (RG) theory either due to their scaling dimension or due to their initial amplitude being of order one \cite{Ady_FMF,PFs_Loss,PF_Mong,vaezi_2}.

We first define standard bosonic fields $\phi_{r \eta \sigma}$ as $R_{\eta \sigma}=e^{i \phi_{1\eta \sigma}}$ and $L_{\eta \sigma}=e^{i \phi_{\bar 1\eta \sigma}}$ with the only non-vanishing commutation relation given by $[\phi_{r\eta \sigma} (x), \phi_{r'\eta' \sigma' } (x')] = i\pi r \de_{rr'} \de_{\eta \eta'} \de_{\sigma \sigma'}  {\rm sgn} (x-x')$. However, the problem is described better in terms of new bosonic fields $\tilde \phi_{r\eta \sigma} = (2 \phi_{r\eta \sigma}  - \phi_{\bar r\eta \sigma})/3$ with $[\tilde \phi_{r\eta \sigma} (x), \tilde \phi_{r'\eta'\sigma'} (x')] = i r(\pi/3)  \de_{rr'} \de_{\eta \eta'} \de_{\sigma \sigma'} {\rm sgn} (x-x').$

The non-quadratic Hamiltonians $H_Z^{ee}$ and $H_t^{ee}$ [see Eqs. (\ref{HZee})-(\ref{Hamflo})] can be expressed in bosonized form as
\begin{align}
H_{Z}^{ee} = 2 g_Z  \sum_{\eta}\cos [3  (\tilde \phi_{\eta \eta \bar 1}-\tilde \phi_{\bar\eta  \eta 1})],\\
H_{t}^{ee} = 2 g_d  \sum_{\eta} \cos [3 (\tilde \phi_{1\eta \eta}-\tilde \phi_{\bar1 \bar \eta \eta})] .
\label{BHflo}
\end{align}
Next, aiming to find bound states, one needs to impose vanishing boundary conditions which is best done by the following unfolding procedure \cite{Ady_FMF,PFs_Loss}. We enlarge the nanowire from $[0, L]$ to $[-L, L]$ and define new fields such that the vanishing boundary conditions are satisfied automatically, 
\begin{align}
 \chi_{\eta \sigma} (x) = \begin{cases}
 \tilde \phi_{\bar{(\eta\sigma)}\eta \sigma} (x), &  x>0 \\
  \tilde \phi_{{(\eta\sigma)} \eta \sigma} (-x)+\pi, &  x<0 \\
\end{cases}.
\end{align}
Next, we define the conjugated fields  $\phi_{1}= \frac{3}{2}\sum_{ \eta \sigma}  \chi_{\eta \sigma}$, $\theta_{1} = \frac{3}{2} \sum_{\eta \sigma} \eta \sigma \chi_{\eta \sigma} $, $\phi_{2}= \frac{1}{2} \sum_{\eta \sigma} \eta \chi_{\eta \sigma}$, and $\theta_{2} = \frac{1}{2} \sum_{\eta \sigma} \sigma \chi_{\eta \sigma}$.

The Hamiltonians take the form
\begin{align}
&H_{Z}^{ee} = 4 g_Z  \cos ( \theta_1) \cos (3 \theta_2) , \ x>0,\\
&H_{t}^{ee} = 4 g_d  \cos (\theta_1) \cos (3 \phi_2) , \ x<0.
\end{align}
To minimize the total energy in the strong coupling regime, the fields get pinned. The first field $ \theta_1$ is uniform over the entire system, $\theta_1=\pi \hat M$, where $\hat M$ is an integer-valued operator. The second field can not be pinned uniformly over the whole system and changes from $\theta_2=\pi (1+\hat M+2\hat l)/3$ for $x>0$ to $\phi_2=\pi (1+\hat M+2\hat n)/3$ for $x<0$, where $\hat l$ and $\hat n$ are integer-valued non-commuting operators with $[\hat n, \hat l]=  3i/4\pi $. The domain wall at $x=0$ hosts a zero-energy parafermion state \cite{Ady_FMF,PFs_Loss} defined by the operator $\alpha_{\pm}$,
\begin{align}
\alpha_{\pm}= e^{i 4 \pi (\hat n\pm \hat l)/3},\  \alpha_{\pm}^3=1.
\end{align}
We note here that coming back to the  time-independent lab frame, the energy of the bound states will stay at zero but the many-body wavefunctions will be periodically changing in time.

\section{Floquet Rashba nanowire proximity-coupled to a superconductor} 
\begin{figure}[t]
\includegraphics[width=0.6\columnwidth]{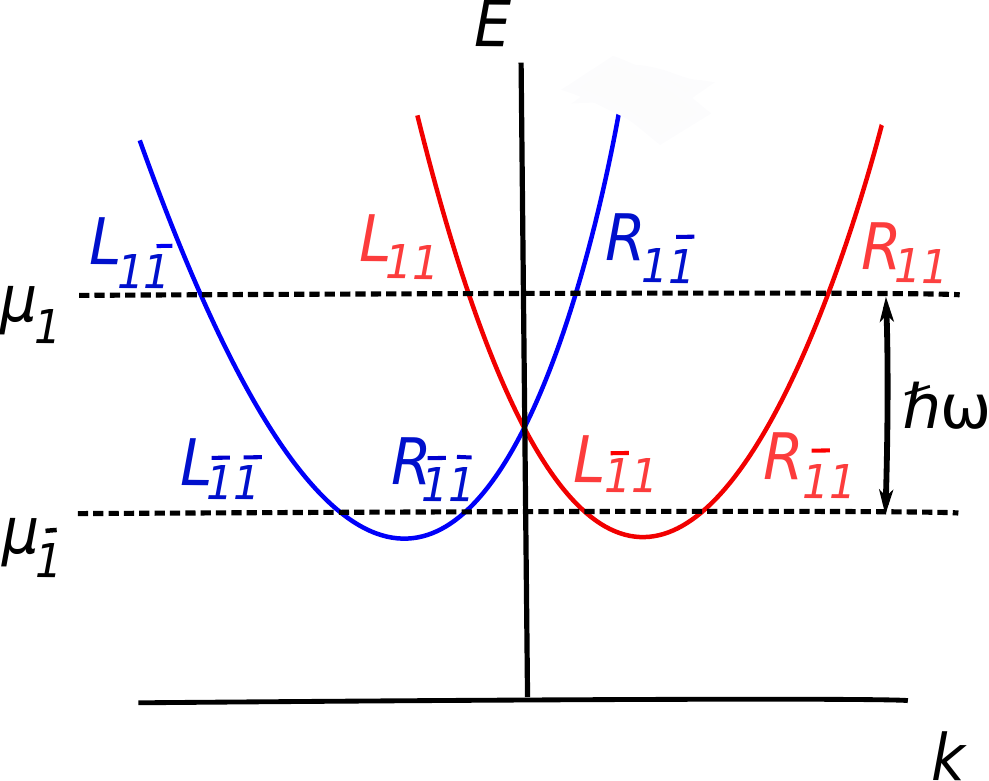}
\caption{The spectrum of the Rashba nanowire used in the second model. The  index $\eta=1$ $(\eta=\bar1)$ is for upper and lower band, $\sigma=1$ $(\sigma=\bar1)$ for spin up (red) [down (blue)].  The chemical potentials $\mu_{1,\bar1}$ and the driving frequency $\omega$ of field $B(t)$ are chosen in such a way that the  smallest Fermi wavevectors of two effective subbands coincide.}
\label{1DS_2s}
\end{figure}

In the second model, we consider a one-band Rashba nanowire proximity-coupled to an $s$-wave superconductor. The system is periodically driven by a time-dependent uniform magnetic field $B(t)$ of amplitude $B_0$ and frequency $\omega$ applied  perpendicular to the SOI vector.
We note that in the first model, the chemical potential was assumed to be close to the SOI energy. However, tuning of the chemical potential gets challenging if the system is coupled to a superconductor. Thus, our second model has an important advantage in that the chemical potential just needs to be below the SOI energy level [see Fig. \ref{1DS_2s}] but does not need to be tuned to a particular value. By adjusting $\omega$ of 
$B(t)$,
one can then tune the Floquet Zeeman  term to be resonant.

The Floquet driving takes place inside the same band. The lower (upper) energy states are labeled by the index $\eta=\bar 1$ ($\eta= 1$) and spin up (down) by $\si=1$ ($\si=\bar 1$). The chemical potential  $\mu_{\bar 1}<0$ lies away from the SOI crossing. The frequency $\omega$ of $B(t)$
is chosen such that $\mu_1 =\hbar \omega + \mu_{\bar 1}$ satisfies the resonance condition both in  energy and momentum space, see Fig. \ref{1DS_2s}. 
The Fermi points in the two bands are given by $k_{F\eta\si \pm} = \si k_{so} \pm k_{so} \sqrt{1+ (\mu_\eta/E_{so})}]$. The driving frequency $\omega $  is determined by the condition $k_{F\bar 1 1 -} =  k_{F  1\bar1 +} $. Again, to characterize the system, we linearize the Hamiltonian density around the Fermi points and keep only slowly varying fields \cite{Composite}. The pairing term becomes
\begin{align}
H_{s}= \sum_{\eta }\De_{sc} [R^\dag_{\eta \bar1} L^\dag_{\eta 1}- R^\dag_{\eta 1} L^\dag_{\eta \bar1}+  \text{H.c.}],
\end{align}
where $\De_{sc}$ is the proximity induced superconducting gap. The  resonant part of the Floquet term takes the form
\begin{align}
H_{d}= t_F \sum_\eta [R_{\eta \bar1}^\dag L_{\bar \eta 1} + \text{H.c.}].
\end{align}
Here, $t_F=g\mu_B B_0$  is the amplitude of the Zeeman coupling in the Floquet representation. 

The corresponding linearized Hamiltonian density is given by
\begin{align}
\mathcal H= \hbar \upsilon_F \hat k \tau_3 + \De_{sc} \tau_1 \sigma_2 \de_2 + \frac{t_F}{2} \eta_1 \de_3 (\tau_1 \sigma_1 + \tau_2 \sigma_2),
\label{H_lin2}
\end{align}
where $\de_{i}$ are the Pauli matrices acting in the electron-hole space. The spectrum of the linearized Hamiltonian is given by $E_{1,\pm}=\pm \sqrt{(\hbar \upsilon_F  k)^2+\De_{sc}^2}$ and $E_{2,\pm}^2= \sqrt{(\hbar \upsilon_F  k)^2+(t_F\pm \De_{sc})^2}$, where $E_{1,\pm}$ is four fold and $E_{2,\pm}$ is two fold degenerate.

If the Floquet process dominates over superconductivity, $0<\De_{sc}<t_F$, the system is in the topological phase and hosts two zero-energy bound states at each of its ends protected by the effective time-reversal symmetry. The corresponding wavefunctions at the left wire end at $x=0$ are given in the basis composed of $\Psi_{\eta\sigma}$ [$(\Psi_{11},\Psi_{1\bar 1}, \Psi_{11}^\dagger,\Psi_{1\bar 1}^\dagger,\Psi_{\bar 11},\Psi_{\bar 1\bar 1},\Psi_{\bar 11}^\dagger,\Psi_{\bar 1\bar 1}^\dagger)$]  by
\begin{align}
&\Phi_{MF1} = (f, i f^*, f^*,-i f, g, i g^*, g^*,-i g)^T, \\
&\Phi_{MF2} = (- i f, -f^*,  i f^*, -f, -i g, -g^*, i g^*,-g)^T, \\
&f= e^{-x/\xi_-} e^{i k_{F\bar 11-} x} -e^{-x/\xi_0} e^{- i k_{F\bar 11+} x},\\
&g= e^{-x/\xi_-} e^{i k_{F 11-} x} -e^{-x/\xi_0} e^{- i k_{F 11+} x}.
\end{align}
Here the localization lengths are given by $\xi_0=\hbar \upsilon_F / \De_{sc}$ and $\xi_-=\hbar \upsilon_F / (t_F-\De_{sc})$.
We note that the two Majorana fermion wavefunctions are connected by an effective time-reversal symmetry transformation, defined as the product of time reversal and band inversion symmetry transformations and given by $U_T=\sigma_2 \tau_1 \eta_3$. Under this symmetry transformation $U_T$ we find $R_{\eta \sigma} \rightarrow (\eta \sigma) L_{\eta \bar{\sigma}}$ and  $L_{\eta \sigma} \rightarrow (\eta \sigma) R_{\eta \bar{\sigma}}$, and thus $U_T^\dag \mathcal{H}^*(-k) U_T= \mathcal{H}(k)$.  We note that, in contrast to Kramers pairs protected by the time-reversal symmetry \cite{Const,bib:Mele2013,bib:Berg2013,Beri,bib:Flensberg2014,bib:Oreg2014,bib:Law2012,bib:Nagaosa2013,bib:Law2014,bib:Tewari2014,bib:Loss12014,bib:Loss22014,bib:Trauzettel2011},  the degeneracy of the pair can be lifted by disorder \cite{Silas,Chen}. Thus, these states are similar to fractional fermions, which similar to MFs possess non-Abelian statistics \cite{Ab} and can be used for quantum computing schemes.

In the presence of strong electron-electron interactions, we repeat the same bosonization procedure as described above (see Sec. \ref{FFM}) for the first model. We find that this setup can also be brought into the fractional topological regime and the many-body ground state consists of ${\mathbb Z}_3$ parafermions, see the Appendix \ref{B}. 

\textit{Conclusions}\textemdash
We proposed two simple one-dimensional setups which host zero-energy modes. In the first setup, we consider a single Rashba nanowire with applied uniform static magnetic field driven by a time-dependent electric field. An important feature of this scheme is that no superconductivity is needed, and thus no restrictions on the magnetic field strengths are required. Due to their intrinsic particle-hole symmetry, promising candidates for this setup are carbon nanotubes \cite{CNT,CNT2,CNT3,CNT4}, graphene \cite{gra,gra2,gra3,gra4,gra5}, and other two-dimensional crystals \cite{sc,mos,mos2,mos3,mos4,mos5,mos6}. For example, the parameter estimates for metallic armchair graphene nanoribbons \cite{gra4} are ($k_B T, t_F, \Delta_Z^*, E_{so}$)=(10, 20, 50, 100) $\mu$eV, which correspond to $B=0.5$ T (applied say, along the ribbon axis), $\omega=50$ GHz for $\mathcal E \approx 40$ mV/$\mu$m ($d_{cv} \approx 1$ nm) applied transverse and in-plane.
We note that the SOI can be generated by spatially rotating magnetic fields \cite{gra4} or by using functionalized graphene \cite{gra5}. In the second setup, we consider a model relying on superconductivity with the resonant driving achieved by applying a time-dependent magnetic field. The advantage of this one-band setup is the flexibility in the positioning of the chemical potential. This feature is especially valuable for semiconducting nanowires with large $g$-factor and with weak proximity-induced superconductivity \cite{Leo,Marcus}. The periodic driving brings both systems from the trivial to the topological phase.
The systems can be tuned  further from  standard  to  fractional topological phase if strong electron-electron interactions are present, which leads in particular to the emergence of  parafermions. The potential realization of such systems could be also in cold atoms or optical lattices. Relaxation and heating effects \cite{Eliashberg,Glazman} are of general concern in Floquet systems \cite{Mitra,Chamon}. It has been shown, however, that these harmful effects can be suppressed  by adiabatic build-up of the fractional state \cite{Usaj} or by engineered baths \cite{Refael}.

We would like to acknowledge Peter Stano for useful discussions. This work was supported by the Swiss National Science Foundation (SNSF) and NCCR QSIT.

\appendix

\section{Parafermions in Floquet Rashba nanowire with superconductivity \label{B}}
\begin{figure}[t]
\includegraphics[width=0.6\columnwidth]{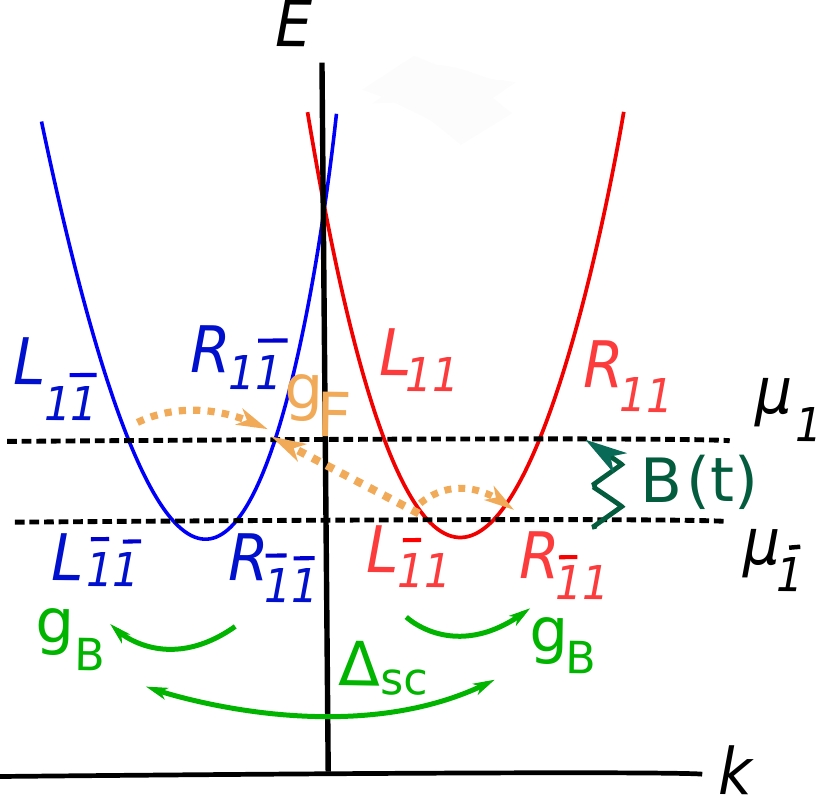}
\caption{The spectrum of the Rashba nanowire modified by the proximity gap $\Delta_{sc}$ and time-dependent magnetic field $B(t)$ in the strong electron-electron interaction regime. The  index $\eta=1$ $(\eta=\bar1)$ is for upper (lower) band, $\sigma=1$ $(\sigma=\bar1)$ for spin up (red) [down (blue)].  The leading term in driving $H_{F}^{ee}$ (yellow arrows) involves two momentum-conserving backscattering terms. The superconductivity term $H_{sc}^{ee}$ (green arrows) commutes with $H_{F}^{ee}$ (orange arrows), therefore they can lead to simultaneous ordering of the corresponding bosonic fields, resulting in the fully gapped energy spectrum with  zero-energy parafermion bound states localized at each wire end.}
\label{1DS_two}
\end{figure}

Similar to the first model considered in the main text, the periodically driven one-dimensional Rashba nanowire proximity-coupled to a superconductor can also be brought into the fractional topological regime. The frequency of the ac magnetic field is chosen to be $\hbar \omega = \mu_1 - \mu_{\bar 1}$, where the chemical potentials are fixed such that $(k_{F\bar1 \bar 1 +}-k_{F\bar1 \bar 1 -}) + (k_{F 11 +}-k_{F 11 -}) =  (k_{F \bar 1 \bar1 +}- k_{F 11 -})$ or $(k_{F\bar1  1 +}-k_{F\bar1 1 -}) + (k_{F 1\bar1 +}-k_{F 1\bar1 -}) =  (k_{F 1 \bar1 +}- k_{F \bar11 -})$  [see Fig. \ref{1DS_two}(b)]. Again, we assume that the driving term $H_{F}^{ee}$ describes the dominant process.

Hence, the leading order term that conserves momentum is given by
 \begin{align}
H_F^{ee}&= g_F  \Big[( R^\dag_{1 \bar1}L_{\bar11})( R^\dagger_{\bar{1} 1}L_{\bar{1}  1}) ( R^\dagger_{1\bar1}L_{1\bar1})\nonumber \\
&\hspace*{1cm} + (R^\dag_{\bar1\bar{1}}L_{1 1})(R^\dag_{\bar1  \bar{1}} L_{\bar1 \bar{1}}) (R^\dag_{1 1} L_{1 1}) + \text{H.c.} \Big].
\end{align}
The superconducting term which commutes with $H_{F}^{ee}$, is given by
\begin{align}
H_{sc}^{ee}&= g_{sc} \Big[( R^\dag_{1 1}L^\dag_{1\bar{1}})( R^\dagger_{1 1}L_{1  1}) ( R_{1 \bar1}L^\dag_{1 \bar{1}}) \nonumber \\
&\hspace*{1cm}+ ( R^\dag_{\bar1 1}L^\dag_{\bar{1}\bar1})( R^\dagger_{\bar1 1}L_{\bar1 1}) ( R_{\bar1 \bar1}L^\dag_{ \bar{1} \bar1})+\text{H.c.} \Big],
\end{align}
where $g_{F}\propto t_F g_{B}^2$ and $g_{sc} \propto\De_{sc} g_{B}^2$. We note that these terms are possible only due to backscattering events of finite strength $g_{B}$. 
We use bosonic fields $\phi_{r \eta \sigma}$ as $R_{\eta \sigma}=e^{i \phi_{1\eta \sigma}}$ and $L_{\eta \sigma}=e^{i \phi_{\bar 1\eta \sigma}}$  with the only non-zero commutation relations given by $[\phi_{r\eta \sigma} (x), \phi_{r'\eta' \sigma' } (x')] = i\pi r \de_{rr'} \de_{\eta \eta'} \de_{\sigma \sigma'}  {\rm sgn} (x-x')$. The problem simplifies by using new fields, therefore introducing $\tilde \phi_{r\eta \sigma} = (2 \phi_{r\eta \sigma}  - \phi_{\bar r\eta \sigma})/3$ with $[\tilde \phi_{r\eta \sigma} (x), \tilde \phi_{r'\eta'\sigma'} (x')] = i r(\pi/3)  \de_{rr'} \de_{\eta \eta'} \de_{\sigma \sigma'} {\rm sgn} (x-x')$.  In terms of the new fields, the non-quadratic Hamiltonian takes the form 
\begin{align}
H_{sc}^{ee} = 2 g_{sc}  \sum_{\eta}\cos [3  (\tilde \phi_{1 \eta  1}+\tilde \phi_{\bar 1  \eta \bar1})], \\
H_{F}^{ee} = 2 g_{F}  \sum_{\eta}\cos [3  (\tilde \phi_{1 \eta \bar1}-\tilde \phi_{\bar 1  \bar\eta 1})].
\end{align}
Again, we double the system size and halve the number of fields in order to satisfy vanishing boundary conditions at the two ends of system \cite{Ady_FMF,PFs_Loss}. The new fields can be written as
\begin{align}
\chi_{1 \eta } (x) = \begin{cases}
 \tilde \phi_{1\eta 1} (x), &  x>0 \\
  \tilde \phi_{\bar{1} \eta 1} (-x)+\pi, &  x<0 \\
\end{cases},\\
\chi_{\bar{1}\eta } (x) = \begin{cases}
 \tilde \phi_{\bar{1} \eta \bar1} (x), &  x>0 \\
  \tilde \phi_{1 \eta \bar1} (-x)+\pi, &  x<0 \\
\end{cases}.
\end{align}

Therefore, the Hamiltonian has the following form
\begin{align}
H^{ee}=\begin{cases}
2 g_{sc}\sum_{\eta}\text{cos}[3(\chi_{1\eta}+\chi_{\bar1 \eta})], &  x>0 \\
2 g_{F}\sum_{\eta}\text{cos}[3(\chi_{1\bar\eta}-\chi_{\bar1 \eta})], &  x<0 \\
\end{cases}.
\end{align}

Next, we transform the chiral fields to conjugate fields $\phi$'s and $\theta$'s as $\chi_{r\eta}=[r\phi_2+ \theta_2+\eta (r\phi_1 +\theta_1)/3]/2$ and get
\begin{align}
H^{ee}=\begin{cases}
4 g_{sc}~\text{cos}(\theta_1) \text{cos}(3\theta_2), &  x>0 \\
4 g_{F}~\text{cos}(\theta_1) \text{cos}(3\phi_2), &  x<0 \\
\end{cases}.
\end{align}
To minimize the energy of the system  \cite{Ady_FMF,PFs_Loss}, we find $\theta_1= \pi \hat M$ (pinned uniformly over the entire wire), $\phi_2= \pi(1+\hat M+2\hat n)/3$  for $x<0$, and $\theta_2= \pi(1+\hat M+2\hat l)/3$ for $x>0$.
Thus, a domain wall is formed between two non-commuting fields, namely $\phi_2$ and $\theta_2$, $[\phi_2(x),\theta_2(x')]=-i\pi/3 ~{\rm sgn}(x-x')$. This gives the non-zero commutator $[\hat n,\hat l]=3i/4\pi$, hence we define two operators which commute with the Hamiltonian and are at zero energy \cite{Ady_FMF,PFs_Loss},
\begin{align}
\alpha_1= e^{i 4 \pi (\hat l-\hat n)/3}; \alpha_{\bar1}= e^{i 4 \pi (\hat l+\hat n)/3}.
\end{align}
These zero energy operators satisfy the parafermionic algebra: $\alpha_1^3=\alpha_{\bar 1}^3=1$ and $\alpha_1 \alpha_{\bar 1}= e^{-2i \pi/3} \alpha_{\bar 1} \alpha_1$. 


\end{document}